\documentclass[12pt]{article}
\usepackage{amsmath}
\usepackage{axodraw4j}
\usepackage{pstricks}
\usepackage{color}
\usepackage{xcolor}
\usepackage{framed}
\usepackage{cite}
\definecolor{shadecolor}{rgb}{0.90,0.90,0.90}
\usepackage{hyperref}
\usepackage{subcaption}
\usepackage[compat=1.1.0]{tikz-feynman}
\usepackage[utf8]{inputenc}

\usepackage{mathtools}
\usepackage{setspace}
\usepackage{amsmath, amssymb, amsthm, float, graphicx}
\numberwithin{equation}{section}
\hypersetup{colorlinks=false, linkcolor=blue, citecolor=red}

\def\beq{\begin{eqnarray}}
\def\eeq{\end{eqnarray}}
\def\be{\begin{equation}}\def\ee{\end{equation}}
\def\nn{\nonumber}
\def\g{\gamma}
\def\r{\rho}
\def\s{\sigma}

\def\a{\alpha}

\def\ex{{\rm{e}}}

\def\D{\Delta}
\def\G{\Gamma}
\def\l{\lambda}

\def\bz{\bar{z}}

\def\la{\langle}
\def\ra{\rangle}

\def\G{\Gamma}

\begin{document}
\title{\bf Correlation functions at the bulk point singularity from the gravitational eikonal S-matrix}
\date{}

\author{Carlos $\text{Cardona}^{\Delta}$\\
${}^{\Delta}$Niels Bohr International Academy and Discovery Center,\\University of Copenhagen, Niels Bohr Institute\\ Blegamsvej 17, DK-2100 Copenhagen Ø, Denmark
}

\maketitle
\vskip 2cm
\abstract{The bulk point singularity limit of conformal correlation functions in Lorentzian signature acts as a microscope to look into local bulk physics in AdS. From it we can extract  flat space scattering processes localized in AdS that ultimate should be related to corresponding observables on the conformal field theory at the boundary. In this paper we use this interesting property to propose a map from flat space s-matrix to conformal correlation functions and try it on perturbative gravitational scattering. In particular, we show that the eikonal limit of gravitation scattering maps to a correlation function of the expected form at the bulk point singularity. We also compute the inverse map recovering a previous proposal in the literature.}

\vfill {\footnotesize 
	carlosgiraldo@nbi.ku.dk}

\newpage

\tableofcontents
%\tableofcontents

\onehalfspacing

\section{Introduction and motivation}
According to the AdS/CFT correspondence \cite{Maldacena:1997re, Witten:1998qj}, correlation functions in the boundary conformal field theory are equivalently described through scattering processes in AdS.  Since the early days of the conjecture a recurrent question has been how to relate the flat space limit of observables in AdS with correlation functions on the boundary's conformal field theory  \cite{Polchinski:1999ry, Susskind:1998vk}.  There are several possible ways to achieve this limit and the convenience of each approach depends on the particular problem one would like to explore. An obvious possibility is to think of AdS as a box of size $R$ that should be rescaled to infinite volume $R\to\infty$\cite{Paulos:2016fap, Paulos:2016but, Fitzpatrick:2011jn}, in this way we also need to rescale energies and spins of the operators inside the correlator to avoid overrun those scales with the infinite radius of AdS.  

One can also use the remarkable similarity between the Mellin representation of conformal correlation functions and flat space s-matrix as a guide line to figure out a proper limit that takes the former into the latter \cite{Penedones:2010ue, Fitzpatrick:2011ia, Fitzpatrick:2011hu, Fitzpatrick:2011dm}. 

A more recent proposal that have gained some attention over the last couple of years to map scattering amplitudes in flat space to conformal correlation functions in lower dimensions, (even though so far is not completely clear how it connects to AdS/CFT, some discussions towards this connection has been undergone recently, see for example \cite{Nandan:2019jas, Hijano:2018nhq,  Hijano:2019qmi}), is motivated by the asymptotic BMS symmetries in gravitational theories
\cite{Bondi:1962px, Sachs:1962zza}, and is based on the observation that the $d$-dimensional conformal symmetry can be linearly realized as a $d+2$-dimensional Lorentz symmetry which in turns allows to write the plane waves in flat Minkowskian space as an expansion in conformal primary wave functions \cite{Pasterski:2016qvg}. For some progress in this direction see \cite{Cheung:2016iub, Ball:2019atb, Cardona:2017keg, Pasterski:2017kqt, Pasterski:2017ylz, Lam:2017ofc, Banerjee:2019tam, Adamo:2019ipt, Stieberger:2018onx, Schreiber:2017jsr, Hijano:2019qmi}

Another approach is to consider scattering processes that under certain conditions are very localized in the middle of AdS and therefore are insensitive to the curvature of the space \cite{Polchinski:1999ry, Susskind:1998vk, Gary:2009ae, Gary:2009mi, Heemskerk:2009pn, Okuda:2010ym}. A suitable Lorentzian configuration to investigate bulk locality is one in which the operators live inside the Milne cone, such as in particular for a four-point correlation function two of the operators insertions are space-like separate and live in the future light-cone of other two space-like separated insertions, as shown in figure \ref{loretnzianconf}. By taking the limit where the Graham determinant of the matrix made up from the insertions distances vanish, the correlation function develops a singularity that pick up the local physics in the bulk. This special singularity has been termed ``The bulk point singularity'' by \cite{Maldacena:2015iua}.  In this paper we follow closely the spirit of \cite{Gary:2009mi,  Heemskerk:2009pn, Maldacena:2015iua} and use this particular method in the case of a four point correlation function.

We can in fact have a more general point of view of the bulk point singularity that might allow us to think in applications for theories without a gravity dual. Four-point correlation functions in conformal field theory are usually represented in an operator product expansion (OPE) throughout conformal blocks. Such representation is very reminiscent of the partial wave expansion of the flat space s-matrix. 
On the other hand, at the bulk point singularity, the conformal blocks are written in terms of spherical harmonic functions, and hence the OPE expansion of the correlation function adopts a similar form as an s-matrix partial wave expansion. In this paper we exploit this similarity to propose a map between these two seemingly unconnected observables.  We use this map to compute Witten diagrams at the bulk point singularity from their simpler flat space Feynman diagrams counterpart. By taking advanced of the all-loop resummation of gravitation scattering in flat space at the eikonal approximation \footnote{The eikonal approximation in AdS spaces have been studied in a series of interesting papers \cite{Cornalba:2006xk, Cornalba:2006xm,  Cornalba:2007zb, Cornalba:2008qf, Cornalba:2007fs}}, we can compute the corresponding resummation of the leading singularity expansion of Witten diagrams \footnote{Recently, the Eikonal approximation in flat space has been used to imposed causality constrains on large$-N$ gauge theories in \cite{Kaplan:2019soo} }.

The remaining of this paper is organized as follows, in section \ref{section2} we review the bulk point singularity limit of a four-point OPE and show how it exposes the associated flat-space scattering. In section \ref{section3} we present a map from the flat-space s-matrix to correlation functions for theories with gravity dual description and test it on the simplest gravitational two-to-two scatterings. In section \ref{section4} we apply the aforementioned map to the resummed eikonal gravitational s-matrix. In section \ref{section5} we compute the inverse map, namely taking a correlation function to a flat s-matrix and recover a previous conjecture made in \cite{Okuda:2010ym}. Section \ref{section6} contains the conclusions.

\section{Bulk point singularity limit of the four-point correlation function}\label{section2}
\begin{figure}[]
\begin{center}
\includegraphics[scale=0.6]{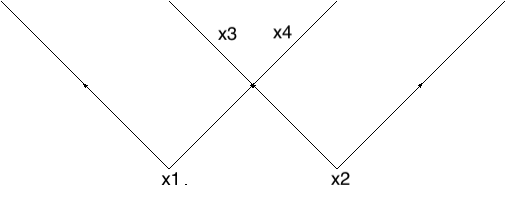}
\end{center}
\caption{\label{loretnzianconf} Lorentzian configuration for the bulk point singularity}
\end{figure}

We want to consider a four-point correlation function of primary scalar operators which for the sake of simplicity we take of having the same conformal dimension. Up to an overall factor, it is a function of cross-ratios only,
\be \label{4p}
\langle {\cal O}_{\varphi}(x_1) {\cal O}_{\varphi}(x_2){\cal O}_{\varphi}(x_3){\cal O}_{\varphi}(x_4)\rangle = 
  \frac{1}{(x_{12}^2)^{\Delta_{\varphi}}(x_{34}^2)^{\D_{\varphi}}}
A(z,\bz)
\ee
where the usual cross-ratios are given by,
\be
z\bz={x_{12}^2x_{34}^2\over x_{13}^2x_{24}^2},\,\quad (1-z)(1-\bz)={x_{14}^2x_{23}^2\over x_{13}^2x_{24}^2}\,.
\ee
However, we will find convenient to use instead  $z=\s \ex^{\rho},~ \bz=\s \ex^{-\rho}$, where $\s$ and $\rho$ can be written in terms of the operator positions as,
\beq
\s^2&=&{x_{12}^2x_{34}^2\over x_{13}^2x_{24}^2}\,, \nn\\
\sinh^2\r&=&{{\rm det}(x_{ij}^2)\over 4 x_{12}^2x_{34}^2x_{13}^2x_{24}^2}\,.
\eeq
We then represent the cross-ratios dependence of the correlator in terms of an S-channel operator product expansion,
\be\label{ope}
A(\r,\s)= 1+\sum_{\D=0}^\infty\sum_{\ell=2n-1} p_{\D,\ell} g_{\Delta,\ell}(\r,\s) ,
\ee
%Let us start considering contributions from double trace operators, $\D_{n,\ell}=2\D_0+2n+\ell+{\g^{(1)}_{n,\ell}\over N^{2}}$
%\be
%f(r,\phi) = 1+\sum_{n=0}^\infty\sum_{\ell=0,2,\dots} p_{n,\ell} g_{\Delta_{n,\ell},\ell}(r,\phi) + {\rm other\ operators},
%\ee
%where ``other operators" are single- and multi-trace operators that contribute at first and higher order in the $1/N^2$ expansion.
We would like to consider the lorentzian configuration depicted at figure \ref{loretnzianconf} and look near the region $\r<<1$. It turns out that in the above given kinematics, the correlator develops  a singularity at $\r=0$\cite{Gary:2009ae} which as been termed as ``the bulk point singularity" by \cite{Maldacena:2015iua}. 
In order to understand the analytic continuation from the Euclidean correlator to the  corresponding Lorentzian configuration, it is useful to use the embedding coordinates  in $\mathbb{ M}^{D+1}$ for euclidean AdS (with $D=d+1$),  
\be\label{boundarypointsEucl}
P^E_{i}=(\cosh\tau_i,\sinh\tau_i,\cos\phi_i,\sin\phi_i,\mathbf{0}^{(d-2)})\,.
\ee
%one can start with the correlator on the Euclidean cylinder with cross-ratios parametrized by,
%\beq
%\ex^{\tau_E+i\phi}&=&{z\over (1+\sqrt{1-z})^2}\,,\nn\\
%\ex^{\tau_E-i\phi}&=&{\bz\over (1+\sqrt{1-\bz})^2}\nn\,,
%\eeq
such that the Lorentzian configuration is obtained by the usual Wick rotation in global time $\tau_E\to i\tau$, taking us to the embedding coordinates  on $\mathbb{ M}^{d+2}$, 
\be\label{boundarypoints}
P_{i}=(\cos\tau_i,\sin\tau_i,\cos\phi_i,\sin\phi_i,\mathbf{0}^{(d-2)})\,.
\ee
In order to maintain convergence of the OPE we should take $\tau\to(\tau-i\xi)$ with $\xi$ small and positive (see \cite{Maldacena:2015iua} for details).
Therefore, the Lorentzian configuration represented in \ref{loretnzianconf} corresponds to points located at,
\beq\label{locations}
 &\tau_{1}=\tau_2=-\pi+i\xi,~~\tau_3=\tau_4=0+i\xi,\nn\\
 &\phi_1=\phi+\pi,\,\phi_2=\phi,\,\phi_3=0,\,\phi_4=\pi\,,
\eeq
The singularity is approached  by taking the limit where $\xi\to 0$ \footnote{See \cite{Gary:2009ae} for details on the analytic continuation from the Euclidean correlator to the bulk-point Lorentzian regime.}.

The cross-ratios in this positions are then given by,
 \beq
\s^2&\sim&{1\over \sin^4\left({\phi\over 2}\right)}\,, \nn\\
\sinh^2\r&\sim& -{\xi^2 \cos^2\left({\phi\over 2}\right)\over \sin^2\left({\phi\over 2}\right)}\,,
\eeq
where we have taken the small $\xi$ limit. 

It is worth to mention the distinction between the configuration considered here, and the so-called light-cone and Regge configurations. In both of the latter cases, distances $x_{13}$ and $x_{24}$ are taken to be time-like while the remaining distances are all space-like. To achieve the Regge limit, we should take on the last configuration $x_1$ going to future null infinite and $x_3$ to past null infinite. This different regions can be reach by moving position on the Lorentzian cylinder $(\tau,\phi)$, as it has been detailed explained in \cite{Maldacena:2015iua}.
For a nice comparison between the analytic continuation from one region to the other, see \cite{Kulaxizi:2017ixa}. 

The parametrization \eqref{locations} provides us with a nice picture for the conditions under which the singularity is developed. In order to satisfy $\r=0$, the condition ${\rm det }({P_i\cdot P_j})=0$ should be satisfied. This happens if the vectors $P_i$ are linearly dependent or if they form a null manifold. 
At the particular locations \eqref{locations} the vectors or ``momentums'' $P_i$ are automatically conserved at $\xi=0$  and are orthogonal to the bulk-point $X=(0,1,\mathbf{0}^{d})$.
However more generally, the matrix  $S_{ij}\equiv{P_i\cdot P_j}$ being singular, has a zero-eigenvector. Following \cite{Maldacena:2015iua}, we can define $n_i$ as the null vector obtained from $P_i$ by removing the second entry such as they are still orthogonal to the bulk point $X=(0,1,\mathbf{0}^{d})$. The entries of the zero-eigenvector can be chosen by
\be
k_a= (-1)^{a-1}{\rm det }'_a(n_{i,b})\,,
\ee 
where the tilde means the determinant is taken by removing column $b$. The zero-eigenvector $k_a$,
\be
\sum_{a}S_{i a}k_a=0\,,
\ee
can be used to choose  a set of conserved momentum by $p_i=k_a n_i$.

Going back to the conformal partial wave expansion \eqref{ope}, at the bulk point singularity regime it has been argued in \cite{Heemskerk:2009pn} that it should be dominated by large values of the exchanged operator conformal dimension. An evidence  of this is given by taking the bulk-point singularity limit of a tree-level Witten diagram and notice that its singularity is stronger than the one for an individual conformal block, therefore the singularity of the given diagram must arise from an infinite sum over conformal blocks and so it would be visible from the tail of the sum at large $\D$ of the analytically continued partial waves. Hence, in order to use the proper OPE representation we need to analytically continue the euclidean  conformal blocks to the lorentzian bulk point kinematics \eqref{locations} and then take the large $\Delta$ limit, while keeping $\xi\D-$finite. This has been explicitly done in \cite{Maldacena:2015iua} and we borrow their result here, 
\be
g_{\Delta,\ell}(\xi,\phi) = 
{\ex^{-i\pi\D}\,2^{1-d\over 2}\ell! \over \sqrt \pi (d-2)_\ell} {C_\ell^{d/2-1}(\cos \phi) \over |\sin\phi|} \sqrt \Delta\, \xi^{3-d\over 2} K_{d-3\over 2}(\Delta \xi) \qquad (\Delta \gg 1, \Delta \xi\ {\rm fixed}).
\ee
Where 
\be
\xi\sim{\r\over \sqrt{1-\s}},\quad\text{and}\quad\sin^2\left({\phi\over 2}\right)={1\over \s}\,.
\ee
%As we can immediately notice, the conformal blocks will be dominated by the bulk point singularity $\xi^{3-d\over 2}$ and therefore the physics gets localized around the boundary configuration \eqref{boundarypoints} which  should comes from a corresponding localized region in the embedded AdS space, therefore allowing us to believe that this region is approximated by the flat space around it.
The conformal block expansion for the four-point function can be then approximated at the leading singularity as,
\be
A(\xi,\phi) = 1+\sum_{\D=0}^\infty\sum_{\ell}\tilde{p}_{\D,\ell} {\ex^{-i\pi\D}\,2^{1-d\over 2}\ell! \over \sqrt \pi (d-2)_\ell} {C_\ell^{d/2-1}(\cos \phi) \over |\sin\phi|} \sqrt \Delta\,  \xi^{3-d\over 2} K_{d-3\over 2}(\Delta  \xi) \,.
\ee
where we have denoted $\tilde{p}_{\D,\ell}$ as the large $\D$ approximation for the OPE coefficients $p_{\D,\ell}$. %Considering that we keep the leading term at the large $\D-$ limit of those coefficients, we can  write the given leading term as,
%\be\label{largdope}
%\tilde{p}_{\D,\ell}=\D^{\alpha(d,\D_{\phi})}\,f(d,\ell)\,.
%\ee
We can see that every individual conformal block develops a  singularity $\xi^{3-d\over 2}$, and therefore  in general we might expect a stronger singularity on the four-point correlator since we have to sum an infinity number of blocks, unless some very precise cancellation occurs, as for example for the case of free conformal field theories, which of course does not develop a bulk point singularity.

We can rewrite the correlator in a more illuminating way as,
\be\label{usualopeatbulkpoint}
A(\xi,\phi) = 1+\sum_{\D=0}^\infty\,  \xi^{3-d\over 2} K_{d-3\over 2}(\Delta  \xi) 
{\sqrt \Delta\over |\sin\phi|}\left(\sum_{\ell}\,{\ex^{-i\pi\D}\tilde{p}_{\D,\ell}\,2^{1-d\over 2}\ell! \over \sqrt \pi (d-2)_\ell} C_\ell^{d/2-1}(\cos \phi)  \right)\,.
\ee
One immediately recognize that the expression in parentheses looks like an $s-$channel $(d+1)-$flat space partial wave expansion of a s-matrix up to an arbitrary global function $h(\D)$. In the following section we will show what this function would be for gravitational dual theories.\footnote{Notice that the Gegenbauer is in $d$ therefore expanding an element in $D=d+1$. Similarly as an S-matrix in $D=3+1$ is expanded by Gegenbauer with $d=3$, i.e, Legendre polinomials. } We therefore define,
\be
 T(\sqrt{s}\equiv\D,\cos{\phi})=h(\D)\sum_{\ell}\, \ex^{-i\pi\D}\tilde{p}_{\D,\ell}\, \widetilde{C}_\ell^{d/2-1}(\cos \phi) \,,
\ee
where we have make the natural identification of the center of mass energy  to the conformal dimension of the operator exchanged $\sqrt{s}=\D$ and have normalized the Gegenbauer polynomials in a more convenient way as,
\be
\widetilde{C}_{\ell}^{d/2-1}(\cos \phi)={2^{1-d\over 2}\ell! \over \sqrt \pi (d-2)_\ell} C_\ell^{d/2-1}(\cos \phi)  \,.
\ee
We therefore conjecture the following relation between the flat-space S-matrix and the associated correlation function at the bulk boundary point,
\be
A^d(\xi,\phi) = 1+\sum_{\D=0}^\infty\,  \xi^{3-d\over 2} K_{d-3\over 2}(\Delta  \xi) {\sqrt \Delta\over |\sin\phi|}\,
T^{d+1}\left(\sqrt{s}=\D,\cos{\phi}\right)\,.
\ee
or more conveniently,
\be\label{conjecture}
\tilde{A}^d(\xi,\phi) = \int_0^\infty d\D\,\,  \xi^{3-d\over 2} K_{d-3\over 2}(\Delta  \xi) {\sqrt \Delta\over |\sin\phi|}\,
T^{d+1}\left(\sqrt{s}=\D,\cos{\phi}\right)\,,
\ee
where the identity has been absorbed in the redefinition $\tilde{A}^d(\r,\phi)$ and given the fact that the sum is dominated by large values of $\D$ we have replaced it by an integration. 

Notice that up to this point we have not made any assumption on the nature of the exchanged operators, other that they have large dimension. Perhaps the most subtle question is if we indeed can interpret the coefficients  $ \ex^{-i\pi\D}\tilde{p}_{\D,\ell}$ as the corresponding partial wave coefficients for physically sensible flat space s-matrix. If that is the case, we should be able to  write it as,
\be\label{doubletwistphase}
 \ex^{-i\pi\D}\tilde{p}_{\D,\ell}=(\ex^{i\delta_{\ell}(\D)}-1)\,.
\ee
 where $\delta_{\ell}$ is the phase shift, which we are taken as complex such as the imaginary part corresponds to the inelasticity. 
\section{Mapping for gravitational dual theories}\label{section3}
In this section we would like to go back to  \eqref{usualopeatbulkpoint} and specialize it to large $N$ theories with gravity dual in order to explore the bulk point singularity limit for correlations functions corresponding to Witten diagrams by using flat-space s-matrix.

As in  \cite{Maldacena:2015iua, Heemskerk:2009pn} we will assume that the main contribution to the bulk point limit of this correlators comes from the exchanging of double twist-operators ${\cal O}_{n,l}\equiv {\cal O}_{\varphi}\partial^{2n}\partial_{\mu_1}\cdots\partial_{\mu_{\ell}}{\cal O}_{\varphi}$ with classical dimension $\D^{\rm class}_{n,\ell}=2\D_{\varphi}+2n+\ell$ and which are present in the operators spectrum  of any conformal field theory in $d>2$ \cite{Komargodski:2012ek, Fitzpatrick:2012yx}. The resulting computation will prove that this is indeed the case. The OPE coefficients for those double-twist operators, have the following form \cite{Heemskerk:2009pn, Fitzpatrick:2010zm}
\beq
p_{\D,\ell}&=& \frac{32 \pi ^{3/2} \Gamma (\Delta -1) (h+l-1) \Gamma (2 h+l-2) \Gamma \left(\frac{l+\Delta }{2}\right)
   \Gamma \left(\frac{1}{2} (-2 h-l+\Delta +2)\right)}{\Gamma
   \left(h-\frac{1}{2}\right) \Gamma (l+1) \Gamma \left(\D_{\varphi}\right){}^2 \Gamma (\Delta -h)
   \Gamma \left(\frac{1}{2} (l+\Delta -1)\right) \Gamma \left(-h+\D_{\varphi}+1\right){}^2}\nn\\
&\times&   \frac{ \Gamma \left(\frac{1}{2} (-4 h-l+\Delta +2)+\Delta
   _{\phi }\right) \Gamma \left(\frac{1}{2} (-2 h+l+\Delta )+\D_{\varphi}\right)}{ \Gamma
   \left(\frac{1}{2} (-2 h-l+\Delta +1)\right) \Gamma \left(\frac{1}{2} \left(-l+\Delta -2 \Delta _{\phi
   }+2\right)\right) \Gamma \left(\frac{1}{2} \left(2 h+l+\Delta -2 \D_{\varphi}\right)\right)}\,,
\eeq
which for large $\D$ can be approximated as (with $2h=d$) \footnote{For large$-N$ conformal field theories, the dimension of the double twist is corrected by a small anomalous dimension  $\gamma_{n}/N$. The large contribution to the dimension comes from large  $n$ and so we are considering $\D\sim n>>\D_{\phi},\,\ell$}
\be
\tilde{p}_{\D,\ell}=\D^{4 \D_{\varphi}-3h} \left(\frac{16^{h-\D_{\varphi}+1} (h+\ell-1) \pi
   ^{3/2} \Gamma (2 h+\ell-2)}{\Gamma \left(h-\frac{1}{2}\right) \Gamma (\ell+1) \Gamma \left(\Delta _{\phi
   }\right){}^2 \Gamma \left(-h+\D_{\varphi}+1\right){}^2}\right)\,. 
\ee
We can then rewrite the OPE as,
\beq A(\xi,\phi) &=& 1+{\cal N}_a\sum_{\D=0}^\infty\,  \xi^{3-d\over 2} K_{d-3\over 2}(\Delta  \xi){\sqrt \Delta\over |\sin\phi|}\D^{4 \D_{\varphi}-3h}\nn\\
&&~~~\times\sum_{\ell}\, \ex^{-i\pi\D}\frac{ (h+\ell-1) \Gamma (2 h+\ell-2)}{ \ell!}\widetilde{C}_\ell^{d/2-1}(\cos \phi) \,,
\eeq
with 
\be
{\cal N}_a= {16^{h-\D_{\varphi}+1}\pi^{3/2}\over\Gamma \left(h-\frac{1}{2}\right)\Gamma \left(\Delta _{\varphi
   }\right){}^2 \Gamma \left(-h+\D_{\varphi}+1\right){}^2}\,.
\ee
So for large $N$ theories we adopt the following modified map, \footnote{Notice that this relation is different from the one proposed in \cite{Gary:2009ae} where the correlation function is integrated against some localized kernels. }
\be\label{conjecturegravity}
 \tilde{A}(\xi,\phi) ={\cal N}_a\int_{0}^\infty d\D \, \xi^{3-d\over 2} K_{d-3\over 2}(\Delta  \xi){\sqrt \Delta\over |\sin\phi|}\D^{4 \D_{\varphi}-3h}\,\D^{d-3}\,T^{(d+1)}(\sqrt s=\D,\phi) \,,
\ee
with 
\be\label{partialwavegravitational}
 T^{(d+1)}(\sqrt s=\D,\phi)=\D^{3-d}\sum_{\ell}\, \ex^{-i\pi\D}\frac{ (h+\ell-1) \Gamma (2 h+\ell-2)}{ \ell!}\widetilde{C}_\ell^{d/2-1}(\cos \phi)\,.
\ee
The reason why we need a specialized map for the case of theories with gravity dual is that in such case double trace operators contribute at  large $N$   at the same order as single trace operators and they are actually the main contribution near the bulk point singularity, whereas for generic theories (meaning theories without a large $N-$expansion) we don't expect double twist operators contributions to dominate over single trace operators at large $\D$ \footnote{Excluding free theories and special supersymmetric theories with susy-protected double trace operators}.
\subsection{Witten diagrams at the bulk point}
In what follows, we would like to consider very simple flat space s-matrices and check if the map proposed above lead us to the expected results. 

The simplest diagram is the one associated to a quartic contact term $T^{(d+1)}=\l$, being $\l$ the coupling at the vertex. 
%It is well know that this diagrams will be dominated by the anomalous dimension for double-twist operators at large $n$, which is given by \cite{Maldacena:2015iua, Heemskerk:2009pn},
%\be\label{anomalouslarge}
%\gamma_{n}\sim{(\D^{d-3} \over 2^{d+2}\pi^{d/2}\Gamma(d/2)}\,,
%\ee
Putting this into the proposed map  \eqref{conjecturegravity} yields,
\be \tilde{A}(\xi,\phi) =  {{\cal N}_a \l\over |\sin\phi|}\int_{0}^\infty d\D\,(\D)^{4 \Delta_{\varphi} -4}\,  (\D\xi)^{3-d\over 2} K_{d-3\over 2}(\D  \xi)\,.
\ee
The type of integral we have to perform has the form,
\beq\label{Ixi}
 I(\xi,\phi) &=& \int_{0}^{\infty}dx\,x^{4   \Delta _{\varphi } -4+2\a}\, (x\, \xi)^{3-d\over 2} K_{d-3\over 2}(x  \xi)\nn\\
&=&\xi ^{-2 \alpha -4 \Delta _{\varphi }+3} \Gamma \left(\alpha +2 \Delta _{\varphi }-\frac{3}{2}\right)
   2^{\frac{1}{2} \left(4 \alpha -d+8 \Delta _{\varphi }-7\right)} \Gamma \left(-\frac{d}{2}+\alpha +2
   \Delta _{\varphi }\right)\,,\nn\\ 
\eeq
leading us to,
\beq \tilde{A}(\r,\phi) &=&{{\cal N}_a\l  2^{\frac{1}{2} \left( -d+8 \Delta _{\varphi }-7\right)}\over |\sin\phi|}{ \Gamma \left( 2\D_{ \varphi} -\frac{3}{2}\right) \Gamma \left(
   2 \D_{ \varphi} -\frac{d}{2}\right)\over \xi ^{4\D_{\varphi} -3}}\nn\\
   &=&-i{{\cal N}_a\l  2^{\frac{1}{2} \left( -d+8 \Delta _{\varphi }-5\right)}\Gamma \left( 2\D_{ \varphi} -\frac{d}{2}\right)}{ \Gamma \left( 2\D_{ \varphi} -\frac{3}{2}\right) \s (1-\s)^{2\D_{\varphi}-2} \over \r^{4\D_{\varphi} -3}}\,.
 \eeq
 where we have used 
 \be
 \sin\phi=2{\sqrt{\s-1}\over \s} 
 \ee
Which up to an unimportant normalization factor is exactly the behavior of the contact Witten diagram in $AdS_{d+1}$ at the bulk point singularity, as has been shown in  \cite{Gary:2009ae}.
\subsubsection{Scalar exchange}
As we have discussed in section \ref{section2}, the bulk point regime explores large energies $\D>>\D_{\varphi}$ \footnote{Even thought we are taken the center of mass energy large, it should be  smaller than the characteristic energy where string effects start being relevant}, therefore we can approximate the exchange propagator to be massless. 
A scalar exchange in flat space is given by,
\be
T^{d+1}={\l\over -t}={\l\over \D^2 \sin^2{\phi\over2}}\,,
\ee
where we have used,
\be\label{tsangle}
-{t\over s}= {q^2\over s}=\sin^2{\phi\over 2}\,.
\ee
Putting it back into \eqref{conjecturegravity}, lead us to an integration \eqref{Ixi} with $\a=-1$ resulting in,
\beq \tilde{A}(\r,\phi) &=&{{\cal N}_a\l  2^{\frac{1}{2} \left( -d+8 \Delta _{\varphi }-11\right)}\over |\sin\phi|\sin^2{\phi\over2}}{ \Gamma \left( 2\D_{ \varphi} -\frac{5}{2}\right) \Gamma \left(
   2 \D_{ \varphi} -\frac{d}{2}-1\right)\over \xi ^{4\D_{\varphi} -5}}\nn\\
   &=&{{\cal N}_a\l  2^{\frac{1}{2} \left( -d+8 \Delta _{\varphi }-11\right)}\Gamma \left(
   2 \D_{ \varphi} -\frac{d}{2}-1\right)\over \sin^2{\phi\over2}}{ \Gamma \left( 2\D_{ \varphi} -\frac{5}{2}\right) \s (1-\s)^{2\D_{\varphi}-4} \over \r^{4\D_{\varphi} -5}}\,,
 \eeq
which again is the expected behavior of an scalar exchange Witten diagram \cite{DHoker:1999kzh, DHoker:1999mqo} at  the bulk point singularity limit.
\subsubsection{Graviton exchange}
\begin{figure}[]
\begin{center}
\includegraphics[scale=0.6]{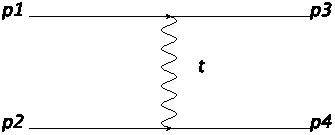}
\end{center}
\caption{\label{gravitonexcha} t-graviton exchanged between two scalars}
\end{figure}

Now consider two energetic scalar particles scattering gravitationally in flat space as in figure \ref{gravitonexcha}.

We are going to borrow the expressions from \cite{KoemansCollado:2019ggb} (see details there and references therein). In the case of four-scalars of equal mass $m$ interacting gravitationaly through  a single graviton exchange we have,
  \be
 T^{d+1}=2i \kappa_D^2{\g(s)\over t}\,,
\ee
with
\be
 \g(s)={(s-2m^2)^2\over 2}-{2\over d-1}m^4\,,
\ee
 and $\kappa_D^2=8\pi {\rm G}_{d+1}$. In the large $s-$limit,
  \be\label{flatexchange}
 T^{d+1}\sim i \kappa_D^2{s^2\over t}\,,
\ee
This expression can as well be easily computed by gluing two three-vertex tree level amplitudes connected by a massless propagator. Namely, for large $s$
there is an effective vertex $\kappa_D s $ and the $t-$channel graviton propagator gives a contribution of ${1\over t}$ as shown in figure \ref{gravitonexcha}.  
By using \eqref{tsangle} the scattering can be rewritten in terms of the energy as,
\be\label{tsangleD}
 T^{d+1}=i \kappa_D^2{\D^4\over \D^2\sin^2{\phi\over 2}}=i \kappa_D^2{\D^2\over\sin^2{\phi\over 2}}\,,
\ee
Putting this into \eqref{conjecturegravity} we have,
\be \tilde{A}(\xi,\phi) =  {{\cal N}_a \l\over |\sin\phi|}\int_{0}^\infty d\D\,(\D)^{4 \Delta_{\varphi} -2}\,  (\D\xi)^{3-d\over 2} K_{d-3\over 2}(\D  \xi)\,.
\ee
leading us to,
\beq A(\r,\phi) &=& \kappa_{d+1}^2{\cal N}_a{1\over {|\sin\phi|\sin^2{\phi\over2}}} { \Gamma \left(2 \Delta _{\varphi
   }-\frac{1}{2} \right) \Gamma \left(-\frac{d}{2}+2 \D_{\varphi}+1\right)\over \xi ^{4 \Delta _{\varphi }-1}} \nn\\
   &=& \kappa_{d+1}^2{\cal N}_a{\Gamma \left(-\frac{d}{2}+2 \D_{\varphi}+1\right)\over\sin^2{\phi\over2}}{ \Gamma \left(2 \Delta _{\varphi
   }-\frac{1}{2} \right) \s(1-\s)^{2\D_{\varphi}-1}\over \r^{4 \Delta _{\varphi }-1} }\nn\,.
\eeq
The corresponding correlator associated to the Witten diagram for a  graviton exchange minimally coupled to the external scalars has been computed in \cite{DHoker:1999kzh}. 
\beq \label{gravresult}
I_{\rm grav}& =&
\left( \frac{6}{\pi^2} \right)^4
\left[ 16\, x_{24}^2\left({1\over 2s}-1\right)\,
D_{4455}+{64\over 9}{x_{24}^2\over x_{13}^2}{1\over s}\, D_{3355}+
{16\over 3}{x_{24}^2\over x_{13}^4}{1\over s}\, D_{2255}
\right. \\
&&\left.+18\,D_{4444}-
{46\over 9\,x_{13}^2}\,D_{3344}-{40\over 9\,x_{13}^4}\,D_{2244} -{8\over 3\,x_{13}^6}\,D_{1144} \right]
 \,,
\eeq
where 
\be 
s \equiv
\frac{1}{2} \frac{x_{13}^2 x_{24}^2}{x_{12}^2 x_{34}^2 + x_{14}^2
x_{23}^2}.
\ee
or equivalently,
\be
x_{34}^2\left({1\over 2s}-1\right)=4{\cos^2{\phi\over 2}\over \sin^2{\phi\over 2}}\,.%-\sinh^2\xi
\ee
The dominant contribution at the bulk point singularity, will come from the D-function $D_{4455}$, which behave as (see appendix \cite{Gary:2009ae})
\be
D_{\D_{\varphi}\D_{\varphi}\D_{\varphi}+1\D_{\varphi}+1}\sim\Gamma \left(2 \Delta _{\varphi
   }-\frac{1}{2} \right){\s(1-\s)^{2\D_{\varphi}-1}\over \r^{4\D_2-1}}
\ee
giving us the following behaviour for the leading bulk-point singularity of the supergravity computation,
\be
I_{\rm grav} \sim{1\over \sin^2{\phi\over 2}}\Gamma \left(2 \Delta _{\varphi
   }-\frac{1}{2} \right){\s(1-\s)^{2\D_{\varphi}-1}\over \r^{4\D_{\varphi}-1}}\,.
\ee
and which up to a normalization factor,  match the result obtained from the flat space s-matrix mapping.
\subsubsection{One loop box}
\begin{figure}[]
\begin{center}
\includegraphics[scale=0.6]{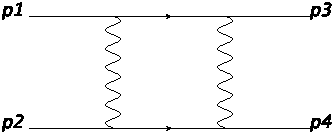}
\end{center}
\caption{\label{gravitonexchabox} t-two graviton  exchanged between two scalars}
\end{figure}

Encouraged by the observations above, we push forward and see if the proposed map translates the perturbative loop expansion in the flat-space scattering to the corresponding perturbative expansion of Witten diagrams at the bulk point limit. For that we now consider the box one-loop contribution at the eikonal limit as shown in figure \ref{gravitonexchabox}, again borrowing the results from  \cite{KoemansCollado:2019ggb}.
\be
T^{d+1}_{\Box}(s,t)= -{\pi^{d+5\over2}\over  (2\pi)^{d+1} }{4 \kappa_{d+1}^4 \g^2(s)\over \sqrt{(s-2m^2)^2-4m^4}}{\Gamma\left({d-3\over2}\right)^2\Gamma\left({5-d\over2}\right)\over \Gamma(d-3) }t^{d-5\over2}
\ee
which in the large $s$ limit can be taken as,
\beq
T^{d+1}_{\Box}(s,t)&=& -{(-1)^{d-5\over2}\pi^{d+5\over2}\over  (2\pi)^{d+1} } \kappa_{d+1}^4 s^3{\Gamma\left({d-3\over2}\right)^2\Gamma\left({5-d\over2}\right)\over \Gamma(d-3) }t^{d-5\over2}=% -{\pi^{d+5\over2}\,\kappa_{d+1}^4\over  (2\pi)^{d+1} } {\Gamma\left({d-3\over2}\right)^2\Gamma\left({5-d\over2}\right)\over \Gamma(d-3) }s^{d+1\over2}\sin^{d-5}{\phi\over 2}
\nn\\
&=& -{\pi^{d+5\over2}\,\kappa_{d+1}^4\over  (2\pi)^{d+1} } {\Gamma\left({d-3\over2}\right)^2\Gamma\left({5-d\over2}\right)\over \Gamma(d-3) }\D^{d+1}\sin^{d-5}{\phi\over 2}\,.
\eeq
putting this back into \eqref{conjecturegravity} we get
\beq \tilde{A}(\xi,\phi) &=& - {{\cal N}_a \over |\sin\phi|\sin^{5-d}{\phi\over 2}}{\pi^{d+5\over2}\,\kappa_{d+1}^4\over  (2\pi)^{d+1} } {\Gamma\left({d-3\over2}\right)^2\Gamma\left({5-d\over2}\right)\over \Gamma(d-3) }\nn\\
&&\times  2^{\frac{1}{2} \left(4d +8 \Delta _{\varphi }-4\right)}{\Gamma \left(d+1 +2 \Delta _{\varphi }-\frac{3}{2}\right)
  \Gamma \left(\frac{d+1}{2} +2
   \Delta _{\varphi }\right)\over \xi ^{4 \Delta _{\varphi }-1+2d} }\nn\\
   &=&-{{\cal N}_a \over \sin^{5-d}{\phi\over 2}}{\pi^{d+5\over2}\,\kappa_{d+1}^4\over  (2\pi)^{d+1} } {\Gamma\left({d-3\over2}\right)^2\Gamma\left({5-d\over2}\right)\over \Gamma(d-3) } 2^{\frac{1}{2} \left(4d +8 \Delta _{\varphi }-4\right)} \Gamma \left(\frac{d+1}{2} +2
   \Delta _{\varphi }\right)\nn\\
&&\times\Gamma \left(d+1 +2 \Delta _{\varphi }-\frac{3}{2}\right) {\s(1-\s)^{2\D_{\varphi}-1+d}
 \over \r ^{4 \Delta _{\varphi }-1+2d} }\,.
\eeq
At this point we don't know the exact answer from the gravity dual computation, so we can think of this result
 a prediction for the bulk point singularity limit of a one loop Witten diagram. The naive interpretation of this is that the exponent for the singularity is given by the same as the single-exchange plus a $2d$ coming from the loop integration.
\section{Eikonal resummation}\label{section4}
By noticing that the bulk point singularity increases with the number of loops, we might wonder if the expansion in loops at the singularity converges. We want to ask the question if similarly as in flat space, the sum of loop Witten diagrams in the bulk point limit ``exponentiate". In order to study that, we want to use the eikonal approximation of the gravitational flat space s-matrix to compute the corresponding bulk point limit of the correlator. 

It is worth to mention here that we will be using the eikonal approximation only as a computable all-loop example of a flat space S-matrix and match it to a corresponding perturbative expansion around the bulk-point region, but we are not trying to relate  it with the eikonal approximation in AdS, which is a quite different limit from the bulk point limit and has been considered in great detail previously at \cite{Cornalba:2006xk, Cornalba:2006xm,  Cornalba:2007zb, Cornalba:2008qf, Cornalba:2007fs}.
\subsection{Eikonal approximation for the gravitational s-matrix}
In this subsection we would like to remind ourselves very quickly the eikonal approximation in gravitation scattering. 
The eikonal regime, namely $-t/s<<1$, turns out to be well described by summing ladder and crossed ladder diagrams, which essentially exponentiates a single-graviton exchange in the so-called phase-shift and in terms of this phase the scattering amplitude is given by the transversal Fourier transform of the resummed graphs, concretely \cite{Levy:1969cr, Amati:1988tn, Amati:1987wq,Muzinich:1987in},
\be\label{eikonalamplitude}
i T_{\rm eik}(s,t)=2s\int d^{d-1}\mathbf{ b}\,\ex^{-i \mathbf{ q}_t\cdot \mathbf{ b}} (\ex^{i\chi(b)}-1)\,,
\ee
Where we denoted by $\mathbf{b}$  the impact parameter and by $\chi$ the phase-shift. This approximation is expected to be valid for relatively large impact parameter withing the following energy region \cite{Giddings:2007qq},
\be
 ({\rm G}_{d+1}E^2)^{1\over d-3}>b>({\rm G}_{d+1}E)^{1\over d-2}\,,
\ee
or equivalently ,
\be
 ({\rm G}_{d+1}\D^2)^{1\over d-3}>b>({\rm G}_{d+1}\D)^{1\over d-2}\,.
\ee
and small angles ,
\be 
{-t\over s}=\sin^2{\phi\over 2}\ll 1\,.
\ee 
${\rm G}_{d+1}$ being the $(d+1)-$ dimensional Newton's gravitational constant. As we mentioned, the leading contribution to the phase-shift at the Eikonal limit is given by the tree level single-graviton exchange  \eqref{flatexchange}.
 %We have borrowed the expression from the recent paper \cite{KoemansCollado:2019ggb}, where the eikonal gravitational scattering was used to extract the two- body classical scattering angle between the two black holes up to the second post-Minkowskian order. In the large $s-$limit we can just write,
%  \be\label{flatexchange} A_{flat}\sim2i \kappa_D^2{s^2\over t}\,.\ee
From this expression, the phase-shift $\chi$ can be computed by inverting the transverse Fourier transform at equation \eqref{eikonalamplitude} and replacing $T_{\rm eik}$ with \eqref{flatexchange}, resulting in,
\be\label{leadeik}
\chi(b)={\kappa_D^2\over 2(D-4)\Omega_{D-3}}{s\over b^{D-4}}\,,
\ee
with $D=d+1$.
One can notice that the phase shift depends only on the magnitude of the transversal impact parameter  vector $\mathbf{b}$ and henceforth when plugging in back into \eqref{eikonalamplitude} we can immediately integrate over angular coordinates, to obtain \cite{Muzinich:1987in},
 \be\label{teik}
i T_{\rm eik}(s,t)=-{2is(2\pi)^{(d-1)/2}\over q_t^{d-3\over 2}}\int_{0}^{\infty}db\,b^{d-1\over 2}J_{d-3\over 2}(q_t\,b)\,(\ex^{i h s/b^{d-3}}-1)\,,
\ee
where $J_{\nu}(x)$ is the Bessel function of first kind and we have defined $h={\kappa_D^2\over 2(D-4)\Omega_{D-3}}$ to unclutter the notation.
In $d=4$ it simplifies further to,
 \be
i T_{\rm eik}(s,t)=-{2is(2\pi)^{3/2}\over q_t^{1\over 2}}\sqrt{{2\over\pi}}\int_{0}^{\infty}db\,b \sin(q_t\,b)\,(\ex^{i h s/b}-1)\,,
\ee
ignoring for now the identity contribution, which will give us the vacuum contribution,  we can integrate the exponential,
% \be
%i T_{\rm eik}(s,t)=-{2is(2\pi)^{3/2}\over q_t^{1\over 2}}{2\over\pi}\frac{2 i  \Delta ^2 h \ker _2\left(2 \sqrt{-i h
%   \Delta ^2} \sqrt[4]{\Delta ^2 \sin ^2\left(\frac{\phi
 %  }{2}\right)}\right)}{\left(\Delta  \sin \left(\frac{\phi
 %  }{2}\right)\right)^{3/2}}\,,
%\ee
%or
 \beq\label{compaceik4d}
i T_{\rm eik}(s,t)%&=&{8(2\pi)^{3/2}h\over\pi}\frac{ \Delta ^2 }{ \sin \left(\frac{\phi}{2}\right)^{2}}\ker _2\left(2\D \sqrt{-i h\D\sin\left(\frac{\phi}{2} \right)}\right)\nn\\&=&
={8(2\pi)^{3/2}h\over\pi}\frac{ \Delta ^2 }{ \sin \left(\frac{\phi}{2}\right)^{2}}\ker _2(\sqrt{2}(1-i)\Theta )\,,
\eeq
 $\ker_{\nu}$ is the Kelvin function defined in terms of the K-Bessel function as,
\be
\ker_{\nu}(z)={\rm Re}\left(\ex^{-\nu\pi i\over 2}K_{\nu}(z\ex^{\pi i\over 4})\right)\,. 
\ee
and we have defined 
\be
\Theta=\D \sqrt{ h\D\sin\left(\frac{\phi}{2} \right)}\,.
\ee
In order to reproduce the leading exchanged graviton  we should take the argument of the Kelvin function small, i.e $\Theta<1$, and therefore the Kelvin function can be Taylor expanded. The leading term of this expansion is just $1/2$ and we recover the single graviton exchange result \eqref{tsangleD}.

The eikonal resummation allow us to take the opposite limit as well. For large values of the argument, the Kelvin function behaves as,
\be 
\ker_{\nu}(z)\sim\sqrt{\pi\over 2 x}\ex^{-x\over \sqrt{2}}\left(f(x)\cos\left({x\over \sqrt{2}}+{\pi\over 8}\right)+g(x)\sin\left({x\over \sqrt{2}}+{\pi\over 8}\right)\right)
\ee
where $f(x)$ and $g(x)$ are given as a series expansion in $1/x$ and whose leading order term is respectively,
\be
f(x)\sim1,\quad{\rm and}\quad g(x)\sim{4\nu^2-1\over 8x\sqrt{2}}\,.
\ee
Therefore, for large argument we can approximate,
\be 
\ker_{\nu}(x)\sim\sqrt{\pi\over 2 x}\ex^{-x\over \sqrt{2}}\cos\left({x\over \sqrt{2}}+{\pi\over 8}\right)
\ee
and the eikonal s-matrix simplifies in this limit to,
 \beq\label{eik4dlargD}
i T_{\rm eik}(s,t)
%&=&-{8(2\pi)^{3/2}h\over\pi}\frac{ \Delta ^2 }{ \sin \left(\frac{\phi}{2}\right)^{2}}\frac{1}{4}(1+i) \sqrt{\frac{\pi }{2}}e^{2 i \eta } \sqrt{\frac{1}{\eta }}\nn\\&=&
   =-2(2\pi)^{3/2}h\frac{ \Delta ^2(i-1) }{ \sin \left(\frac{\phi}{2}\right)^{2}} H^{(1)}_{1\over2}(2\Theta)\,,
\eeq
being $H^{(1)}_{\nu}(2\Theta)$ the Hankel function of the first kind. Interestingly the amplitude behaves oscillatory for large $\Theta$, but is important to notice that in this ultra energetic regime, we expect quantum gravity contributions to start dominating and therefore we should not expect \eqref{eik4dlargD} to hold.
 
At this point is worth looking into the regime of validity of the parameters. The eikonal approximation of the s-matrix is valid as long as the impact parameter is larger than the Schwarzschild radius of a mass corresponding to the center of mass energy  $b>R_{S}(\D)\sim(G_D \D)^{1/(D-3)}$, in other words $\Theta<b^{d-2}$, and as we just mentioned in the paragraph above, $\Theta$ can not be too large.

On the other hand, we also have to satisfy $\xi \D\sim {\cal O}(1)$ and therefore,
  \be\label{constrain1}
  %{\xi b^{(D-3)} \over G_{D}}>1\,\,\rightarrow \,\,
  \xi b^{(d-2)}>G_{D}
  \ee
%   So 
 %  \be
%     {\xi^2 b^{(d-3)} \over G_{D}}= {\xi b^{(d-2)} \over G_{D}}\left({\xi\over b}\right) 
%   \ee
This means that we should consider a large impact parameter compare to $\xi$, but at the same time we need a large AdS radius such as $b<<R_{AdS}$, in order to avoid contributions from the curvature of space time, so the conformal field theory should be strongly coupled.
\subsection{From eikonal s-matrix to the bulk point singularity}
We now want to see how the correlation function looks like at the bulk point singularity as computed from the eikonal flat space result.
For that we need to plug \eqref{teik} into \eqref{conjecturegravity},

\beq
 \tilde{A}(\xi,\phi) &=&  {2i(2\pi)^{(d-1)/2}{\cal N}_a \over |\sin\phi|}\int_{0}^\infty d\D\,(\D)^{4 \Delta_{\varphi} -4}\,  (\D\xi)^{3-d\over 2} K_{d-3\over 2}(\D  \xi)
{\D^2\over \D^{d-3\over 2} \sin^{d-3\over 2}{\phi\over 2}}\nn\\
&&\qquad\qquad~~~~~~\times\int_{0}^{\infty}db\,b^{d-1\over 2}J_{d-3\over 2}(\D\sin{\phi\over 2}\,b)\,(\ex^{h \D^2/b^{d-3}}-1)\,.
\eeq
This integral is hard to compute in general dimension, but we can restrict ourselves again to $d=4$, yielding a fairly large expression, 

\beq
 \tilde{A}(\xi,\phi) &=& {(2\pi)^{3/2}{\cal N}_a \over |\sin\phi|\sin^{1\over 2}{\phi\over 2}}\frac{ h\, u(\phi)^{-\frac{2 \Delta _{\varphi }}{3}-\frac{1}{2}}}{180\, \xi  \sin
   ^{\frac{3}{2}}\left(\frac{\phi }{2}\right)}\times\nn\\
   && \left(\Gamma \left(\frac{4 \Delta _{\varphi }}{3}\right) \Gamma
   \left(\frac{4 \Delta _{\varphi }}{3}+2\right) \, _4F_5\left(\frac{2
   \Delta _{\varphi }}{3}+\frac{1}{2},\frac{2 \Delta _{\varphi
   }}{3}+1,\frac{2 \Delta _{\varphi }}{3}+\frac{3}{2},\frac{2 \Delta
   _{\varphi
   }}{3};\frac{7}{6},\frac{4}{3},\frac{3}{2},\frac{5}{3},\frac{11}{6};\eta\right)\right.\nn\\
   &&\left.\times
   \sin \left(\frac{2 \pi  \Delta _{\varphi }}{3}\right) \xi ^5+5 \cos
   \left(\frac{1}{3} \left(2 \pi  \Delta _{\varphi }+\pi \right)\right)
   \Gamma \left(\frac{4 \Delta _{\varphi }}{3}-\frac{1}{3}\right)
   \Gamma \left(\frac{4 \Delta _{\varphi }}{3}+\frac{5}{3}\right) \right.\nn\\
   &&\left.\times
   _4F_5\left(\frac{2 \Delta _{\varphi }}{3}-\frac{1}{6},\frac{2 \Delta
   _{\varphi }}{3}+\frac{1}{3},\frac{2 \Delta _{\varphi
   }}{3}+\frac{5}{6},\frac{2 \Delta _{\varphi
   }}{3}+\frac{4}{3};\frac{5}{6},\frac{7}{6},\frac{4}{3},\frac{3}{2},\frac{5}{3};\eta\right) \sqrt[6]{u(\phi)} \xi
   ^4\right.\nn\\
   &&\left.
   +20 \left(-\cos \left(\frac{1}{6} \left(4 \pi  \Delta _{\varphi
   }+\pi \right)\right) \Gamma \left(\frac{4}{3} \left(\Delta _{\varphi
   }+1\right)\right) \Gamma \left(\frac{4 \Delta _{\varphi
   }}{3}-\frac{2}{3}\right) \right.\right.\nn\\
   &&\left.\left.\times _4F_5\left(\frac{2 \Delta _{\varphi
   }}{3}-\frac{1}{3},\frac{2 \Delta _{\varphi }}{3}+\frac{1}{6},\frac{2
   \Delta _{\varphi }}{3}+\frac{2}{3},\frac{2 \Delta _{\varphi
   }}{3}+\frac{7}{6};\frac{2}{3},\frac{5}{6},\frac{7}{6},\frac{4}{3},\frac{3}{2};\eta\right) \sqrt[3]{u(\phi)} \xi
   ^3\right.\right.\nn\\
   &&\left.\left.+3 \cos \left(\frac{2 \pi  \Delta _{\varphi }}{3}\right) \Gamma
   \left(\frac{4 \Delta _{\varphi }}{3}-1\right) \Gamma \left(\frac{4
   \Delta _{\varphi }}{3}+1\right) \right.\right.\nn\\
   &&\left.\left. \times _4F_5\left(\frac{2 \Delta
   _{\varphi }}{3}-\frac{1}{2},\frac{2 \Delta _{\varphi
   }}{3}+\frac{1}{2},\frac{2 \Delta _{\varphi }}{3}+1,\frac{2 \Delta
   _{\varphi
   }}{3};\frac{1}{2},\frac{2}{3},\frac{5}{6},\frac{7}{6},\frac{4}{3};\eta\right)
   \sqrt{u(\phi)} \xi ^2 \right.\right.\nn\\
   &&\left.\left. 
   -6 \Gamma
   \left(\frac{4}{3} \left(\Delta _{\varphi }-1\right)\right) \Gamma
   \left(\frac{4 \Delta _{\varphi }}{3}+\frac{2}{3}\right) \,
  \right.\right.\nn\\
   &&\left.\left.\times _4F_5\left(\frac{2 \Delta _{\varphi }}{3}-\frac{2}{3},\frac{2 \Delta
   _{\varphi }}{3}-\frac{1}{6},\frac{2 \Delta _{\varphi
   }}{3}+\frac{1}{3},\frac{2 \Delta _{\varphi
   }}{3}+\frac{5}{6};\frac{1}{3},\frac{1}{2},\frac{2}{3},\frac{5}{6},\frac{7}{6};\eta\right) \right.\right.\nn\\
   &&\left.\left.\times\left(u(\phi)\right)^{2/3} \sin \left(\frac{1}{3} \left(2 \pi  \Delta
   _{\varphi }+\pi \right)\right) \xi \right.\right.\nn\\
   &&\left.\left.+6 \Gamma \left(\frac{4 \Delta
   _{\varphi }}{3}-\frac{5}{3}\right) \Gamma \left(\frac{4 \Delta
   _{\varphi }}{3}+\frac{1}{3}\right)\right.\right.\nn\\
   &&\left.\left.\times \, _4F_5\left(\frac{2 \Delta
   _{\varphi }}{3}-\frac{5}{6},\frac{2 \Delta _{\varphi
   }}{3}-\frac{1}{3},\frac{2 \Delta _{\varphi }}{3}+\frac{1}{6},\frac{2
   \Delta _{\varphi
   }}{3}+\frac{2}{3};\frac{1}{6},\frac{1}{3},\frac{1}{2},\frac{2}{3},\frac{5}{6};\eta\right) \right.\right.\nn\\
   &&\left.\left.\left(u(\phi)\right)^{5/6} \sin \left(\frac{1}{6} \left(4 \pi  \Delta
   _{\varphi }+\pi \right)\right)\right)\right)
\eeq
Where $\eta\equiv\frac{\xi ^6 \csc ^2\left(\frac{\phi }{2}\right)}{2916 h^2}$ and $u(\phi)\equiv-h^2\sin^2\left(\frac{\phi}{2}\right)$.
This is not a very illuminating expression, but we can see right away from the global factor at the first line that the leading singularity behaves as $1/\xi^{d-3}$, and the other terms are polynomials in $\xi$ smoothing out the singularity. Also the argument $\eta$ in the hypergeometric functions is very small so all the hypergeometrics are essentially one. The function $u(\phi)$ is also very small. The leading term is then given by,
\beq
 \tilde{A}(\xi,\phi) &=&i {(2\pi)^{3/2}{\cal N}_a \over |\sin\phi|}\frac{ 2 h^{1-4\D_{\varphi}\over3}\, \sin
   \left(\frac{\phi }{2}\right)^{\frac{2-4 \Delta _{\varphi }}{3}} \sin \left(\frac{1}{6} \left(4 \pi  \Delta _{\varphi }+\pi
   \right)\right) \Gamma \left(\frac{4 \Delta _{\varphi
   }-5}{3}\right) \Gamma \left(\frac{4 \Delta
   _{\varphi }+1}{3}\right)}{3\, \xi  \sin
   ^2\left(\frac{\phi }{2}\right)} \,.\nn\\
\eeq 
We believe this correlator should be thought as a resummation of loop ladder diagrams at the bulk point singularity of Witten diagrams with only graviton exchanges.

The reason why we expect a singularity of the form $1/\xi^{d-3}$ for the full correlator has been explained in \cite{Maldacena:2015iua}, and it comes from the fact that at the $\xi\to 0$ limit the configuration \eqref{locations} is invariant under transformations by the group $SO(1,d-2)$, and therefore the singularity should arises from the action of this non-compact group. Given the above argument is natural to believe that this singularity behavior will be present in any interacting conformal field theory. Another evidence is that the singularity is present at the level of individual conformal blocks and unless very precise cancelations occurs at the OPE expansion we might expect the singularity to survive. Let us consider quickly the case for the $O(N)$ model as a simple example of this.

The four-point function of the $O(N)$ model, has been computed at leading order in a large $N$ expansion in \cite{Lang:1991kp} (see also \cite{Alday:2015ewa} for more recent analysis of anomalous dimension in the $O(N)$ model) and is given by,
\be\label{4sigmas}
\la\s_a(x_1)\s_b(x_2)\s_c(x_3)\s_c(x_4)\ra={\eta_1h\G(h) \over (h-2)\G(h-1)^2 N }{f_{abcd}(u,v)\over (x_{12}^2x_{34}^2)^{\D_{\s}}}\,,
\ee
with
\be
f_{abcd}(u,v)=\delta_{ab}\delta_{cd}\bar{D}_{1,h-1,1,h-1}(u,v)+\delta_{ad}\delta_{bc}\bar{D}_{1,h-1,h-1,1}(u,v)+\delta_{ac}\delta_{bd}\bar{D}_{h-1,h-1,1,1}(u,v)\,,
\ee
and $2h=d$. We then can use the results from Appendix B \cite{Gary:2009ae} to extract the small $\rho$ behavior of the $D-$functions as we did in previous sections,
 \be
\bar{D}_{\D_1,\D_2,\D_1,\D_2}(u,v)\sim-2\pi^{3/2}\G\left(\D_1+\D_2-{3\over 2}\right){\s(1-\s)^{\D_1+\D_2-2}\over (\rho^2)^{\D_1+\D_2-{3\over 2}}}\,.
\ee
The first  term of the four-point correlation function \eqref{4sigmas} at the small $\rho$ limit, looks like,
\beq\label{4sigmasr}
&&\la\s_a(x_1)\s_b(x_2)\s_c(x_3)\s_c(x_4)\ra=\nn\\
&=&-{\eta_1h\G(h) \G\left(h-{3\over 2}\right)\over (h-2)\G(h-1)^2 N }{2\pi^{3/2}\over (x_{12}^2x_{34}^2)^{\D_{\s}}}{\s(1-\s)^{h-2}\over (\rho^2)^{h-{3\over 2}}}=\nn\\
&=&-{\eta_1h\G(h) \G\left(h-{3\over 2}\right)\over (h-2)\G(h-1)^2 N }{2\pi^{3/2}\over (x_{12}^2x_{34}^2)^{\D_{\s}}}{1\over |\sin\phi|\xi^{d-3}}\,,
\eeq
%comparing this with the result from the integration \eqref{Ixi} we can conclude that one possible associated flat s-matrix for the $O(N)$ model should have a phase shift scaling as 
%\be \delta_{\ell}= {1\over N}\D^{2d-3-4\D_{\s}}= {1\over N}\D^{1-2d}\,.\ee
displaying a singularity of the form $1/\xi^{d-3}$ as claimed.
 \section{Inverse map}\label{section5}
In this section we would like to invert the formula \eqref{conjecturegravity} in order to find a representation for the flat space s-matrix in terms of conformal correlation functions.  In the following we are going to focus in  $d=4$, but we believe the arguments can be straightforwardly generalized to other dimensions.
 
 In order to invert equation \eqref{conjecturegravity}  we are going to use the following integral,
\be 
\int_{0}^{\infty}d\xi\, \xi K_{1\over 2}(A\,\xi) K_{1\over 2}(-B\,\xi)=-{\pi\over 2(A-B)\sqrt{-AB}}\,,
\ee 
allowing us to write \eqref{conjecture} as,
 \be
{\cal N}_a\int_{0}^{\infty}d\xi\,\xi^{3\over 2} K_{1\over 2}(-B\,\xi)\tilde{A}^{(4)}(\xi,\phi) = -{\pi\over 2}\int_0^\infty d\D  {\D^{4\D_{\varphi}-{d\over2}-3}\over |\sin\phi|}\,
{T^{d+1}\left(\sqrt{s}=\D,\cos{\phi}\right)\over (\D-B)\sqrt{-B}}\,.
\ee
If we consider a polynomially bounded S-matrix, this integral is peaked around the region $ \D\sim B$, and therefore we can approximate it as,
 \be
T^{d+1}\left(\sqrt{s}=\D,\cos{\phi}\right)\sim{2i{\cal N}_a\over\pi}{|\sin\phi|\, \D^{{d+7\over2}-4\D_{\varphi}}} \int_{0}^{\infty}d\xi\,\xi^{3\over 2} K_{1\over 2}(-B\,\xi)\tilde{A}^{(4)}(\xi,\phi)\,.
\ee
This expression can in turn be tought as a conformal field theory rederivation of Penedones-Okuda formula derived from the gravity side in \cite{Okuda:2010ym} (see equation (2.13) there)\footnote{In order to compare, notice that in \cite{Okuda:2010ym} equation (2.13), there is a global factor of $l_s^3\sqrt{stu}\sim \D^3 |\sin\phi|$}.
Even though we were unable to proved it for general dimension, we find natural to conjecture the form of this limit at arbitrary dimensions as,
 \beq\label{invertedmap}
T^{d+1}\left(\sqrt{s}=\D,\cos{\phi}\right)\sim{2i{\cal N}_a\over\pi}{|\sin\phi|\, \D^{{d+7\over2}-4\D_{\varphi}}} \int_{0}^{\infty}d\xi\,\xi^{d-1\over 2} K_{d-3\over 2}(-B\,\xi)\tilde{A}^{(d)}(\xi,\phi)\,.
\eeq
In principle, we can plug in any four-point correlation function and associate to it a flat s-matrix throughout the mapping above. However, this does not imply we will get a physically acceptable flat space s-matrix associated to a particular theory. Nevertheless, we can use this as a representation for scattering in flat space as an expansion in conformal invariant functions, regardless whether the latest is indeed a physical observable of a particular conformal field theory.

%\subsection{Transverse space at the bulk point}
%Now we would like to present a representation for the s-matrix as an integral over the transverse hyperbolic space at the future Milne cone.
 
%By using the integral representation of the modified Bessel function,
%\be K_{\nu\over 2}={\sqrt{\pi}\over \Gamma\left({\nu+2\over 2}\right) }\left({z\over 2}\right)^{\nu\over 2}\int_{0}^{\infty}\ex^{-z \cosh(\theta)}\sinh^{\nu}(\theta)d\theta\,.\ee
%We can rewrite \eqref{invertedmap},
% \beq &&T^{d+1}\left(\sqrt{s}=\D,\cos{\phi}\right)\sim{2i{\cal N}_a\over\pi}{|\sin\phi|\, \D^{{d+7\over2}-4\D_{\varphi}}}  \nn\\
%&&~~~~~~\times{\sqrt{\pi}\over \G\left({d-2\over 2}\right)}\left({-\D\over 2}\right)^{d-3\over2}{1\over\Omega_{H_{d-1}}}\int_{H_{d-1}}d^{d-1}{\bs{\xi}}\,\ex^{-\bs{\D}\cdot\bs{\xi}}(A^d(\xi,\phi)-1)\,.
%\eeq
%Where we have used hyperbolic coordinates. In this form, the s-matrix is given by a integral over the 
\section{Conclusions}\label{section6}
In this paper we have studied the bulk point singularity limit of correlation functions in generic conformal field theories.
By using the fact that the conformal blocks at the bulk point singularity can be written in terms of spherical harmonics, we have proposed an expansion of four-point correlation functions in terms of a s-matrix in flat space.

We have use this mapping to transform flat-space gravitational scattering to correlation functions associated to Witten diagrams in AdS at the bulk point singularity, and shown that indeed the map reproduce the known results from direct previous calculations of Witten diagrams. 

We have also shown that the eikonal resummation of gravitational scattering maps to a full correlation function with the expected behavior at the bulk point singularity that might be thought as a resummation of Witten graviton ladder exchanges around the singularity.

Finally, we have computed the inverse map transforming conformal invariant functions into flat space scattering reproducing a previous proposal computed from the gravity side in \cite{Okuda:2010ym}, providing a representation of the s-matrix as an expansion in conformal basis at the bulk singularity.

It would be nice to contrast the computations of this paper against  bulk point singularity limits of higher order loop Witten diagrams perhaps by using any of the growing recent methods on the subject \cite{Cardona:2017tsw, Alday:2018pdi, Alday:2017xua, Aharony:2016dwx, Bertan:2018khc, Yuan:2018qva, Aprile:2017qoy, Giombi:2017hpr} 

The eikonal approximation in AdS space have been thoroughly studied in a series of inspiring papers \cite{Cornalba:2006xk, Cornalba:2006xm,  Cornalba:2007zb, Cornalba:2008qf, Cornalba:2007fs}. It would be interesting to see if there is a similar treatment for the ideas pursued in this paper.
\section*{Acknowledgements}
I would like to thank to Alexander Zhiboedov and Poul H. Damgaard for enlightening discussions. This work is supported in part by the Danish National Research Foundation.
%%%%%%%%%%%%%%%%%%%%%%%%%%%%%%%%%%%%
%%%%%%%%%%%%%%%%%%%%%%%%%%%%%%%%%%%%%
%%%%%%%%%%%%%%%%%%%%%%%%%%%%%%%%%
\appendix

\bibliographystyle{JHEP}
\bibliography{LSSCH}
\end{document}